\documentclass{aa}  

\usepackage{graphicx}
\usepackage{txfonts}
\usepackage{xspace}
\usepackage{comment}
\usepackage{amsmath}

\newcommand{\chandra}{\textit{Chandra}\xspace}
\newcommand{\bootes}{Bo\"otes\xspace}
\newcommand{\source}{J1430\xspace}
\newcommand{\fluxcgs}{erg cm$^{-2}$ s$^{-1}$\xspace}
\newcommand{\lumcgs}{erg s$^{-1}$\xspace}
\usepackage[colorlinks, citecolor={blue}]{hyperref}
\begin{document}

   \title{A new distant Giant Radio Galaxy in the Bo\"otes field serendipitously detected by Chandra}

   \author{Alberto Masini
          \inst{1,2}
          \and
          Annalisa Celotti\inst{1,3,4,5}
          \and 
          Paola Grandi\inst{2}
          \and 
          Emily Moravec\inst{6}
          \and 
          Wendy L. Williams\inst{7}
          }

   \institute{SISSA - International School for Advanced Studies, Via Bonomea 265, 34151 Trieste, Italy \email{amasini@sissa.it} 
   \and 
   INAF - Osservatorio di Astrofisica e Scienza dello Spazio di Bologna, Via Gobetti 93/3, 40129 Bologna, Italy 
   \and
   INAF - Osservatorio Astronomico di Brera, Via E. Bianchi 46, I-23807, Merate, Italy
   \and
   INFN - National Institute for Nuclear Physics, Via Valerio 2, 34127 Trieste, Italy
   \and
   IFPU - Institute  for  Fundamental  Physics  of  the  Universe,  Via  Beirut  2, 34151 Trieste, Italy
   \and
   Astronomical Institute of the Czech Academy of Sciences, Bo\v cn\'i II 1401/1A, 14000 Praha 4, Czech Republic
   \and
   Leiden Observatory, Leiden University, PO Box 9513, NL-2300 RA Leiden, the Netherlands}
   \date{}

 
  \abstract
   {Giant Radio Galaxies (GRGs) are the largest single structures in the Universe. Exhibiting extended radio morphology, their projected sizes range from 0.7 Mpc up to 4.9 Mpc. LOFAR has opened a new window on the discovery and investigation of GRGs and, despite the hundreds that are today known, their main growth catalyst is still debated.}
   {One natural explanation for the exceptional size of GRGs is their old age. In this context, hard X-ray selected GRGs show evidence of restarting activity, with the giant radio lobes being mostly disconnected from the nuclear source, if any. In this paper, we present the serendipitous discovery of a distant ($z=0.629$), medium X-ray selected GRG in the Bo\"otes field.}
   {High-quality, deep \chandra and LOFAR data allow a robust study of the connection between the nucleus and the lobes, at a larger redshift so far inaccessible to coded-mask hard X-ray instruments.}
   {The radio morphology of the GRG presented in this work does not show evidence for restarted activity, and the nuclear radio core spectrum does not appear to be GPS-like. On the other hand, the X-ray properties of the new GRG are perfectly consistent with the ones previously studied with Swift/BAT and INTEGRAL at lower redshift. In particular, the bolometric luminosity measured from the X-ray spectrum is a factor of six larger than the one derived from the radio lobes, although the large uncertainties make them formally consistent at $1\sigma$. Finally, the moderately dense environment around the GRG, traced by the spatial distribution of galaxies, supports recent findings that the growth of GRGs is not primarily driven by underdense environments.}
   {}

   \keywords{galaxies: active}

   \maketitle
%

\section{Introduction}

Giant Radio Galaxies (GRGs) are the biggest single structures in the Universe. They are conventionally defined as extragalactic radio sources whose projected linear size on the sky is larger than 0.7 Mpc. 
\par Although well known for decades \citep[e.g.,][]{willis74}, these spectacular objects have recently significantly grown in number thanks to sensitive low frequency radio surveys, such as the LOFAR \citep{vanhaarlem13} Two-metre Sky Survey \citep[LoTSS;][]{shimwell17}. With hundreds of GRGs now discovered \citep{dabhade20}, the main drivers of their large spatial growth have begun to be investigated on a statistical basis. The reason why GRGs are able to reach such sizes is not yet clear. The current possible physical explanations for the exceptional size of GRGs are (1) they could be very old radio galaxies, giving their lobes enough time to expand in the circum-galactic medium, (2) they could have incredibly efficient and collimated relativistic jets, or (3) perhaps there is a sufficiently low-density environment around the GRG that allows the jets and lobes to propagate and expand almost freely.

\begin{figure*}
   \centering
   \includegraphics[width=0.49\textwidth]{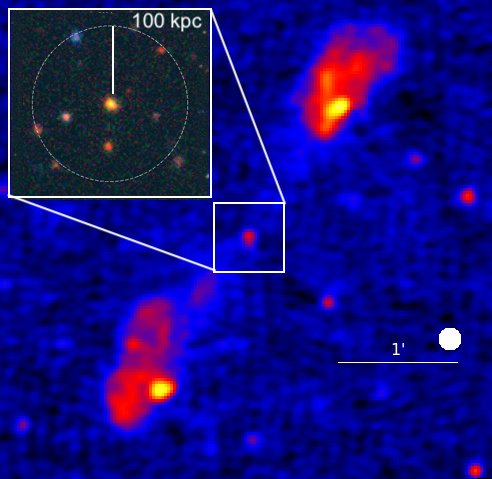}
   \includegraphics[width=0.50\textwidth]{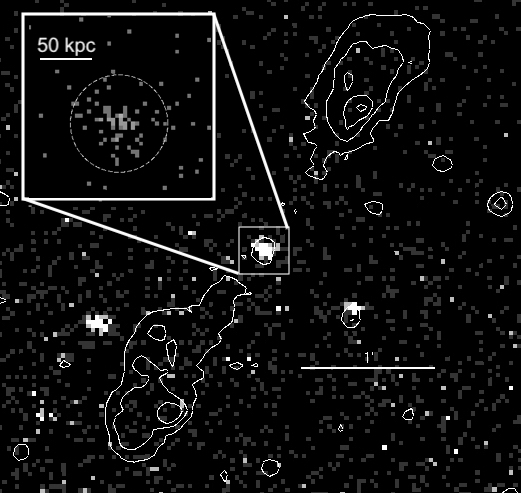}
   
   \includegraphics[width=0.245\textwidth]{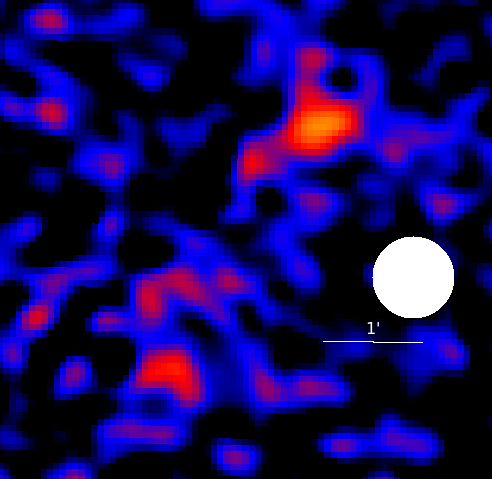}
   \includegraphics[width=0.245\textwidth]{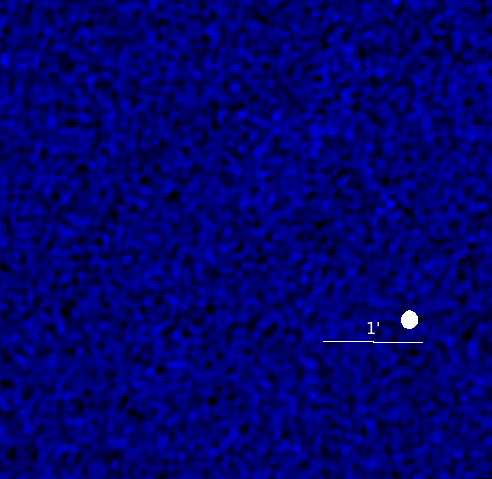}
   \includegraphics[width=0.245\textwidth]{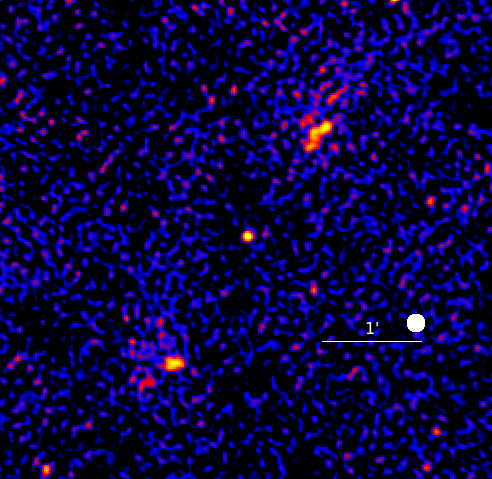}
   \includegraphics[width=0.245\textwidth]{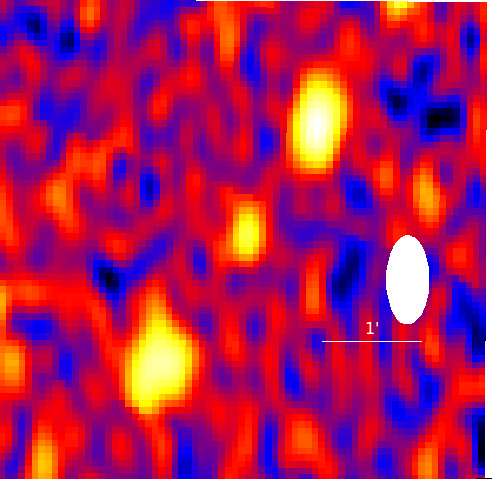}
   
   \caption{\textit{Top left.} LOFAR 150 MHz \citep[][]{tasse20} cutout of the GRG, where the white ellipse labels the beam size of the instrument. The inset shows the optical composite $grz$ DECaLS \citep{decalsdr8} DR8 image, with the white circle marking 100 projected kpc from the galaxy. \textit{Top right.} \chandra 0.5-7 keV \citep{masini20} cutout of the GRG, spatially rebinned with a $4\times 4$ pixel scale, with the white LOFAR contours overlaid. The X-ray image demonstrates that the core is clearly detected, unlike the lobes and hotspots. The inset shows a zoom-in on the core, with a native \chandra pixel scale. The dashed white circle in the inset labels the PSF size encircling $\sim 90\%$ of the counts. Hence, the X-ray emission is consistent with being point-like. \textit{Bottom row.} Radio cutouts. In all the images, the white ellipses label the beam size of the instrument. From left to right, GMRT 150 MHz \citep{williams13}, VLA 325 MHz \citep{coppejans15}, GMRT 608 MHz \citep{coppejans16}, WSRT 1.4 GHz \citep{devries02}. While all the three components (core and lobes/hotspots) are detected at 608 MHz and 1.4 GHz, only the lobes are detected by GMRT at 150 MHz and the whole GRG is undetected by VLA at 325 MHz.}
   \label{fig:grg_images}
\end{figure*}

\par Recent studies have found increasing evidence that the environment of a GRG does not play a significant role in its growth, and that underdense environments are not necessarily required to grow a radio galaxy to giant size; in fact, \citet{dabhade20} found that $\sim 10\%$ of LOFAR GRGs reside in clusters, which are dense environments \citep[see also][]{tang20}; furthermore, \citet{lanprochaska20} explored the position of $\sim 100$ GRGs in the cosmic web and their satellites properties, finding consistency with the results for optically and radio selected control samples matched in luminosity and color. This is in contrast to regular radio galaxies, for which the environment does have an effect on their properties \citep[e.g.,][]{croston19,massaro20,moravec20}. Interestingly, while the environmental factor appears to be negligible, evidence has been recently found that selecting GRGs in the hard X-ray band could provide meaningful insights about their origin. For instance, \citet{ursini18} focused on a sample of local ($z<0.24$) hard X-ray selected GRGs \citep{bassani16,bassani20} finding that their nuclei are powered by radiatively efficient accretion, and that their nuclear X-ray luminosity is larger than the jet power derived from the lobes, hinting to an old lobes/restarted nuclear activity scenario. Such a scenario has been further supported by \citet{bruni20}, that found morphological evidence that the vast majority of hard X-ray selected GRGs could be restarting their nuclear activity, their giant radio lobes being the relics of past accretion episodes.
\par Given these interesting recent results, studying the X-ray properties of GRGs in softer bands becomes important, as the hard X-ray selected ones are intrinsically the brightest of the population at any redshift and may show a different behavior with respect to their lower (X-ray) luminosity counterparts. Furthermore, investigating GRGs in softer X-ray bands could provide a better statistical estimate of their number at higher redshifts.
\par Here, we report the serendipitous discovery of a new distant GRG in the NOAO Deep-Wide Field Survey \bootes field \citep{jannuzi&dey99}. Thanks to field wide area ($\sim 9.3$ deg$^2$) and superb multi-wavelength coverage, it has a great potential for discovery, enhanced by the recent addition of LOFAR observations at 150 MHz \citep{williams16,retanamontenegro18,tasse20} and deep \chandra coverage \citep{masini20}. Despite being previously detected in the shallow \chandra XBOOTES survey with just 5 photons \citep{murray05,kenter05}, the nuclear source (also known as CXOXB J143009.8+351957 - \source hereafter) is now firmly detected with 74 counts over the $0.5-7$ keV band, and coincides with the geometric center of a spectacular radio structure, extended for $\sim 4'$, seen in the LOFAR image at 150 MHz (Figure \ref{fig:grg_images}). Given its spectroscopic redshift \citep[$z=0.629$,][]{kochanek12}, \source's extension reaches $\sim 1.66$ Mpc. As pointed out above, the \chandra detection allows us to probe a lower X-ray luminosity regime inaccessible at the same redshift with hard X-ray coded-mask instruments such as Swift/BAT and INTEGRAL.
\par The paper is structured as follows: in Section \S\ref{sec:radio_prop} the radio properties of the new GRG we discovered are described, while Section \S\ref{sec:xray_prop} focuses on its X-ray spectral properties. In Section \S\ref{sec:lbol} we estimate the bolometric luminosity and the accretion rate of the AGN using different indicators. Section \S\ref{sec:environment} investigates the large scale environment around the GRG. In Sections \S\ref{sec:discussion} and \S\ref{sec:conclusions}, the discussion and summary of our findings are presented. In this work we assume a flat $\Lambda$CDM cosmology with $H_0 = 69.6$ km s$^{-1}$ Mpc$^{-1}$ and $\Omega_{\rm M} = 0.286$.
Unless otherwise stated, uncertainties  are quoted at 68\% of confidence level throughout the paper.

\begin{table}
\caption{Collective information on the source.}
\label{tab:info}
\centering                                    
\begin{tabular}{l c l}          
\hline\hline                        
Parameter & Value & Ref. \\ \hline \\
    R.A. (J2000) & 217.541135 & \\[1mm]
    DEC. (J2000) & 35.332854 & \\[1mm]
    $z$ & 0.629 & (1) \\[1mm]
    Proj. size (Mpc) & $\sim1.66$ & (4) \\[1mm]
    $\log{R_{\rm X}}$ & $\sim -3.6$  & (4) \\[1mm]
    Core dominance & $0.14\pm0.01$ & (4) \\[1mm]
\hline
    $f^{\rm C}_{150, \rm LOFAR}$ & $0.92\pm0.17$ & (4)\tablefootmark{a} \\[1mm]
    $f^{\rm NL}_{150, \rm LOFAR}$ & $30.22\pm0.74$ & (4)\tablefootmark{a} \\[1mm]
    $f^{\rm SL}_{150, \rm LOFAR}$ & $27.66\pm0.65$ & (4)\tablefootmark{a} \\[1mm]
    
    $f^{\rm C}_{150, \rm GMRT}$ & $0.20\pm1.36$ & (4)\tablefootmark{a} \\[1mm]
    $f^{\rm NL}_{150, \rm GMRT}$ & $27.56\pm5.65$ & (4)\tablefootmark{a} \\[1mm]
    $f^{\rm SL}_{150, \rm GMRT}$ & $18.99\pm4.95$ & (4)\tablefootmark{a} \\[1mm]
    
    $f^{\rm C}_{325, \rm VLA}$ & $< 3.06$\tablefootmark{$\dagger$} & (4)\tablefootmark{a} \\[1mm]
    $f^{\rm NL}_{325, \rm VLA}$ & $12.87\pm4.31$ & (4)\tablefootmark{a} \\[1mm]
    $f^{\rm SL}_{325, \rm VLA}$ & $4.72\pm3.78$ & (4)\tablefootmark{a} \\[1mm]
    
    $f^{\rm C}_{608, \rm GMRT}$  & $0.48\pm0.21$ & (4)\tablefootmark{a} \\[1mm]
    $f^{\rm NL}_{608, \rm GMRT}$ & $6.27\pm0.87$ & (4)\tablefootmark{a} \\[1mm]
    $f^{\rm SL}_{608, \rm GMRT}$ & $3.98\pm0.77$ & (4)\tablefootmark{a} \\[1mm]
    
    $f^{\rm C}_{1.4, \rm WSRT}$ & $0.62\pm0.05$ & (2) \\[1mm]
    $f^{\rm NL}_{1.4, \rm WSRT}$ & $2.44\pm0.11$ & (2) \\[1mm]
    $f^{\rm SL}_{1.4, \rm WSRT}$ & $2.15\pm0.10$ & (2) \\[1mm]
    
    $f_{5, \rm VLA}$ & $0.4 \pm 0.4$ & (3)\tablefootmark{b} \\[1mm]
    $f_{31, \rm SZA}$ & $0.90\pm0.16$ & (3) \\[1mm]
    $L_{150, \rm LOFAR}$ (\lumcgs)  & $1.52 \pm 0.03 \times 10^{41}$ & (4)\\
\hline    
    $\Gamma$ & $\sim 2$ & (4)\\
    $N_{\rm H}$ (cm$^{-2}$) & $1.3^{+0.5}_{-0.3} \times 10^{23}$ & (4)\\
    $f_{\rm s}$ (\%) & $4.8^{+3.5}_{-2.5}$ & (4)\\
    $F_{\rm X}$ (\fluxcgs) & $\sim 7.8 \times 10^{-14}$ & (4)\\
    $L_{\rm X}$ (\lumcgs)  & $1.3 \pm 0.5 \times 10^{44}$ & (4)\\
\hline                                             
\end{tabular}
\tablefoot{ All flux densities are given in mJy. The subscripts C, NL, and SL refer to the core, north lobe and south lobe components, respectively.
\tablefoottext{a}{Due to inconsistent radio flux densities presented in the public catalogs at 150 MHz, we measured them again from the radio images.}
\tablefoottext{$\dagger$}{The $3\sigma$ upper limit over the flux density of the core at 325 MHz is reported.}
\tablefoottext{b}{We assumed a 100\% uncertainty on this value since it is missing an uncertainty in the public catalog of \citet{muchovej10}.}
}
\tablebib{
(1)~\citet{kochanek12}; (2) \citet{devries02}; (3) \citet{muchovej10}; (4) This work.
}
\end{table}

\section{The radio properties of \source}\label{sec:radio_prop}

\begin{figure}
   \centering
   \includegraphics[width=0.45\textwidth]{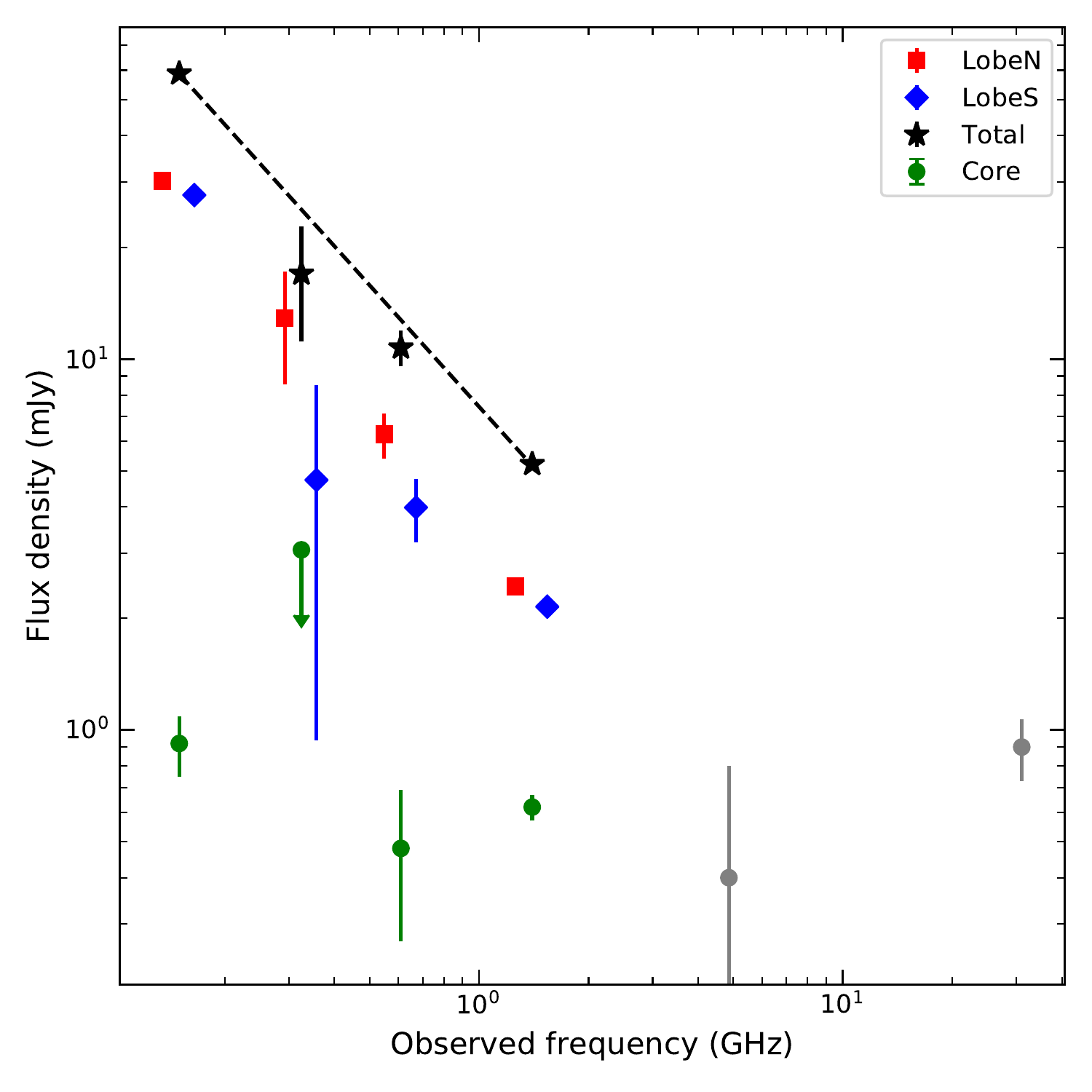}
   \caption{Radio spectrum of the GRG, with flux densities measured at LOFAR  150 MHz \citep{tasse20}, VLA 325 MHz \citep{coppejans15}, GMRT 608 MHz \citep{coppejans16}, WSRT 1.4 GHz \citep{devries02}, VLA 5 GHz \citep{muchovej10}, and SZA 31 GHz \citep{muchovej10}. The green filled circles mark the core data up to 1.4 GHz (the VLA 325 MHz data point being a $3\sigma$ upper limit), while the blue diamonds and red squares refer to the southern and northern lobes, respectively, and have been slightly shifted in frequency to improve the clarity of the plot. The black stars label the total emission from the three components, fit by a power law of slope $\sim-1$ (black dashed line). At higher frequencies (gray circles) the emission is likely dominated by the core, which shows a flat spectrum across the whole range of frequencies.}
   \label{fig:grg_spinfo}
\end{figure}

Recent work has shown that the vast majority of hard X-ray/soft $\gamma$-ray selected GRGs show either restarted radio morphology, such as double double lobes or x-shaped morphology, or Gigahertz-peaked source (GPS) like nuclear radio spectra, which suggest the nuclear emission to be young \citep{bruni20}.
In our case, \source has been instead detected in the medium X-ray band with \chandra. Given the wide frequency coverage of the \bootes field, \source is present in different radio catalogs, from 150 MHz up to 31 GHz. The available radio cutouts are shown in the bottom row of Figure \ref{fig:grg_images}. The GMRT 150 MHz image \citep{williams13} shows a $\sim 3.5 \sigma$ detection of the northern lobe and a possible hint for the southern lobe, but no core emission is detected. The VLA 325 MHz image \citep{coppejans15} shows no detection based on peak flux, but possibly very low surface brightness extended emission. The GMRT 608 MHz image \citep{coppejans16} clearly detects the core and the hotspots of the lobes; the same holds for the WSRT 1.4 GHz image \citep{devries02}.
Recently, new deep LOFAR data on this field have been released \citep{tasse20}, with improved image quality thanks to a direction-dependent calibration and reaching a RMS level of $30~\mu$Jy beam$^{-1}$, the best so far for this field \citep[see Figure 10 of][]{tasse20}. The total integration time of 80 hours has been spread over 8 scans of 10 hours each, covering almost 70 deg$^2$ of area observed at full resolution of $6"\times 6"$.
The exquisite LOFAR image at 150 MHz, shown in the top left panel of Figure \ref{fig:grg_images}, shows prominent lobes with bright hotspots, as usually happens in FRII \citep{fanaroffriley74} radio galaxies. The hotspots almost symmetrically bracket the radio core emission, and there is a hint for the radio jets. 
We notice that there is some radio diffuse emission in the lobes extending well beyond the hotspots ($\sim 332$ kpc and $\sim 166$ kpc beyond the northern and southern hotspots, respectively), which could trace an older population of radio-emitting electrons and could represent the relic of a past emission with very similar jet orientation. Even if the lobes and the hotspots could be associated to two different accretion episodes, their temporal separation (few Myr, assuming the same propagation speed of $\sim 0.1c$) would be much smaller than the age of the last accretion episode. Given its size $D$ and 1.4 GHz power $P$, \source would lie among the other GRGs in the well known $P-D$ diagram. In particular, it would be consistent with the evolution of a $10^{45}$ \lumcgs jet after $\sim 400$ Myr in the framework of \citet{hardcastle18}.
Furthermore, the recessed hotspots could also be due to the bending of the jets along our line of sight if, e.g., \source is moving through some dense environment perpendicularly to the plane of the sky. In summary, because the hotspots are both $> 500$ kpc away from the nucleus already, there is no conclusive morphological evidence for a recently restarted activity.
The extended emission dominates the total radio flux up to at least 1.4 GHz (the core dominance parameter is $0.14\pm0.01$). Its bright low frequency emission results in a radio slope between 150 MHz and 1.4 GHz (defining $f_\nu \propto \nu^\alpha $) $\alpha = -1.09 \pm 0.02$, consistent with the bulk of the GRGs population reported by \citet{dabhade20}. Furthermore, the radio core spectrum is rather flat: a log-linear fit to the core data (green and grey points in Figure \ref{fig:grg_spinfo}) between 150 MHz and 31 GHz, assuming that the 5 and 31 GHz fluxes are dominated by the core and no lobes contribution is present, returns a slope of $\alpha_{\rm Core} = 0.01\pm0.07$.
\par The general radio properties of \source look typical of a normal-sized FRII, and if the redshift were much lower ($z < 0.17$)\footnote{We note that although the spectroscopic redshift of \source is not robust, it is fully consistent with the photometric redshift as derived by \citet{duncan18b}, $z_{\rm phot} = 0.56\pm 0.06$.}, this object would be classified as a local FRII.

\section{The X-ray Properties of \source}\label{sec:xray_prop}

\begin{figure}
   \centering
   \includegraphics[width=0.45\textwidth]{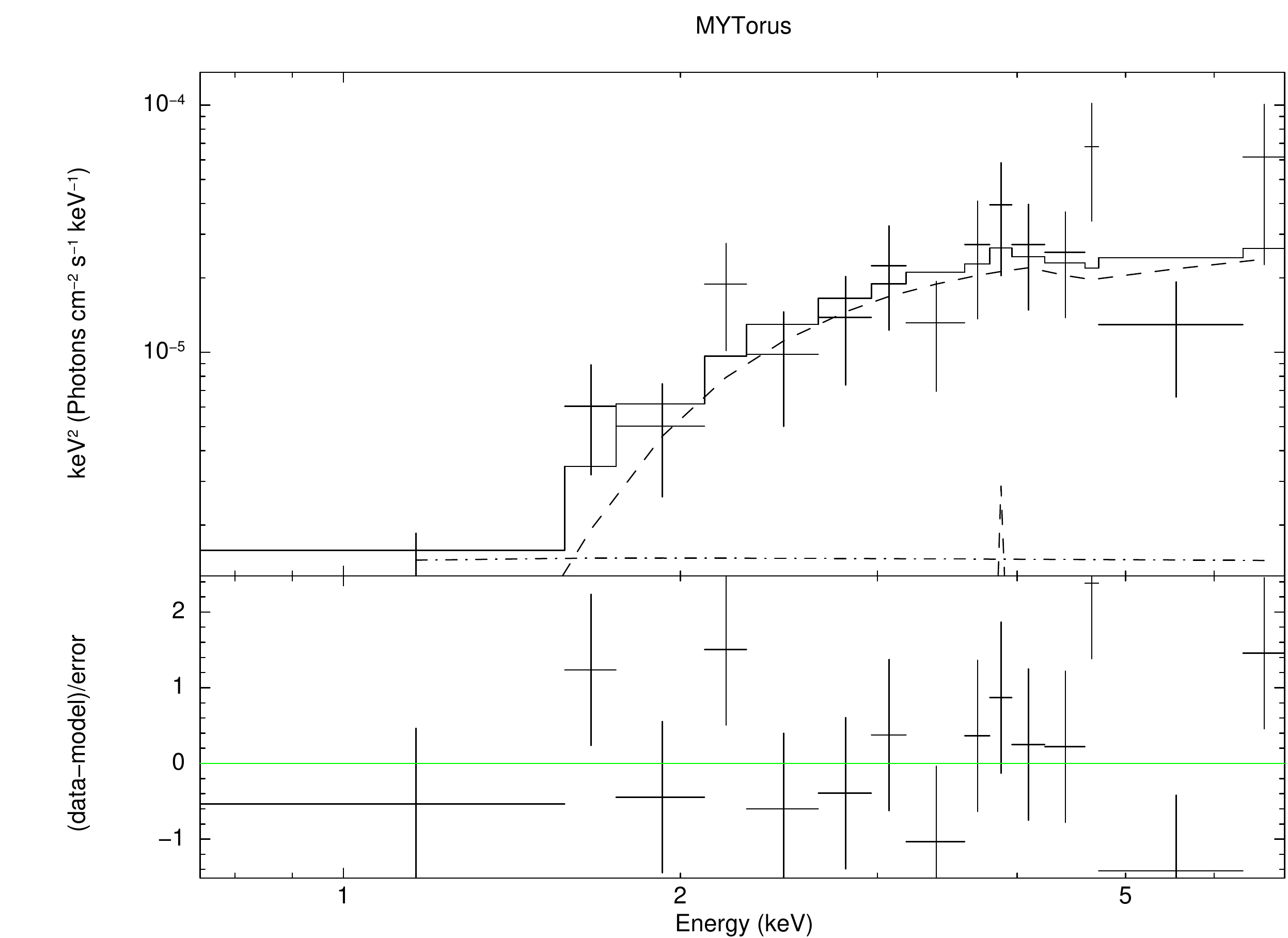}
   \caption{Unfolded spectrum of \source with the best-fit MYTorus model. The solid black line marks the total model, the dashed components are the transmitted and line components of the MYTorus model, while the dot-dashed line labels the soft scattered power law. The curvature of the spectrum below $\sim 5$ keV implies moderate obscuration.}
   \label{fig:xraysp}
\end{figure}

Much information about the physical mechanisms happening in the vicinity of the AGN can be obtained by studying its X-ray spectrum. The core of \source, coincident with the optical position of the host galaxy, is detected as part of the \chandra Deep Wide-Field Survey \citep{masini20} with 74 photons in the $0.5-7$ keV band, corresponding to a significance of $\sim 8\sigma$, while the hotspots and the lobes are undetected in X-rays. The spectrum is extracted with the CIAO \citep{fruscione06} v4.12 task \texttt{specextract} and analyzed with XSPEC \citep{arnaud96} v.12.9.1. The spectrum is rebinned with \texttt{grppha} to have at least one count per bin, and the Cash statistic \citep{cash79} is adopted during the fit. The Galactic column density at the position of the source is $N_{\rm H, Gal} = 1.17 \times 10^{20}$ cm$^{-2}$ \citep{kalberla05}.
\par When fit with a simple power law absorbed by a Galactic column density, the photon index is very hard ($\Gamma = 0.30 \pm 0.25$), indicative of the presence of substantial absorption along the line of sight (or substantial reflection above a few keV). Although the fit is formally acceptable (CSTAT/dof = 53/66), such a hard power law is very atypical; given the fact that \source is an FR II, some degree of obscuration along the line of sight may be expected, and it can reach quite substantial values \citep[e.g., $N_{\rm H} \gtrsim 10^{23}$ cm$^{-2}$;][]{belsole06,evans06}.
\par Thus, a more realistic model including photoelectric absorption and Compton scattering is adopted. We also modeled the possibility that a small fraction of the primary coronal power law is scattered into our line of sight, represented by a soft component arising below a few keV in the spectrum. The parameters of such a scattered power law are fixed to the ones of the primary continuum. The XSPEC implementation of this phenomenological model is the following:

\begin{multline}
\text{Phenom. model} = \overbrace{\texttt{phabs}}^{\text{Galactic $N_{\rm H}$}}\times \{ \overbrace{\texttt{zwabs} \times \texttt{cabs} \times \texttt{zpowerlw}}^{\text{Intrinsic absorbed emission}} + \\ + \underbrace{\texttt{const}\times \texttt{zpowerlw}}_{\text{Soft scattered powerlaw}}\}
\end{multline}

The presence of a scattered power law is required by the data at 98\% confidence level, with a scattered fraction of $f_{\rm s} = 3.2^{+9.0}_{-2.3}\%$, consistently with the values commonly observed in local obscured Seyferts \citep[e.g.,][]{ricci17BASS}. The best-fit (CSTAT/dof = 47/64) photon index is much more reasonable despite the large uncertainty ($\Gamma = 2.0 \pm 0.8$), which reflects in the large uncertainty on the column density ($N_{\rm H} = 1.3^{+0.6}_{-0.5} \times 10^{23}$ cm$^{-2}$). The two parameters are degenerate, with harder slopes coupled to lower absorption. When an AGN is obscured by large amount of gas, a narrow Fe K$\alpha$ line at 6.4 keV is commonly observed due to fluorescence of Fe atoms in the dusty torus. We thus added a narrow Gaussian line at 6.4 keV and tested if the data require its presence, and how significant it would be. Due to the low photon statistics, the line is not statistically significant (CSTAT/dof = 46/63), with an equivalent width (EW) of EW $= 141^{+213}_{-141}$ eV. Similarly, a cold reflection component \citep[i.e., pexrav;][]{pexrav95} is not required by the data, leaving the fit statistic unchanged.
\par To have a better physical description of the X-ray spectrum, we check the results obtained with the previous model with Monte-Carlo based toroidal models commonly adopted in the literature, such as MYTorus \citep{murphy09} and Borus02 \citep{balokovic18}. They both assume a smooth medium (although with different geometry) surrounding a central, isotropic source of X-rays, and self-consistently treat absorption, scattering, and fluorescence. A standard MYTorus model is implemented in XSPEC as follows:

\begin{multline}
\text{MYTorus model} = \overbrace{\texttt{phabs}}^{\text{Galactic $N_{\rm H}$}}\times \{ \overbrace{\texttt{zpowerlw}\times \texttt{MYTZ}}^{\text{Intrinsic emission}} + \\ 
+ \underbrace{\texttt{const}\times \texttt{MYTS}  + \texttt{const}\times \texttt{MYTL}}_{\text{Reprocessed emission}} + \underbrace{\texttt{const}\times \texttt{zpowerlw}}_{\text{Soft scattered powerlaw}}\}
\end{multline}

A fit with this model (CSTAT/dof = 47/64) returns a photon index $\Gamma=2.0$, although unconstrained (the fit is largely insensitive to its exact value), with $N_{\rm H} = 1.3^{+0.5}_{-0.3} \times 10^{23}$ cm$^{-2}$, and a soft power law scattered fraction of $f_{\rm s} = 4.8^{+3.5}_{-2.5}\%$. The best-fit MYTorus model is shown in Figure \ref{fig:xraysp}. Perfectly consistent results are obtained with the Borus02 model, and confirm the picture provided earlier by the obscured power law model.
\par In summary, \source is heavily obscured with a photon index that, despite unconstrained, has a best-fit value consistent with the general, radio quiet, AGN population. A very useful quantity that can be directly measured from an accurate X-ray spectral analysis is the intrinsic (i.e. corrected for absorption) X-ray luminosity of the source. The MYTorus model returns an intrinsic $2-10$ keV flux of $F_{\rm X} \sim 7.8 \times 10^{-14}$ \fluxcgs, implying a rest-frame, intrinsic $2-10$ keV luminosity of $L_{\rm X} = 1.3 \pm 0.5 \times 10^{44}$ \lumcgs. The uncertainties over the luminosity are computed from the joint photon index and power law normalization uncertainties, following \citet{boorman16} and \citet{masini19}. The measured X-ray luminosity of \source is perfectly consistent with the other hard X-ray selected GRGs \citep{ursini18}: the higher sensitivity of \chandra allowed to detect and study a similar object at a larger redshift. The results shown above confirm that the nucleus of \source is actively accreting; the accretion rate can be constrained combining the bolometric luminosity of the system and the black hole mass. We derive them both in the following Section.

\subsection{Radio loudness and jet component in the X-ray spectrum}\label{subsec:rljet}

Given the measured column density, we expect the optical light of the AGN to be heavily extincted (this is indeed confirmed by the results of Section \S \ref{subsec:sed}); thus, the most suitable radio loudness parameter in this case is the one defined by \citet{terashimawilson03}, which exploits the radio to X-ray ratio: $\log{R_{\rm X}}=\nu L_{\nu, 5 \rm GHz}/L_{\rm X}$, with a radio-loud threshold set at $\log{R_{\rm X}}\gtrsim -4.5$. We use the rest frame, de-absorbed $2-10$ keV X-ray luminosity to compute the radio loudness, and interpolate the flat radio core spectrum at the observed frequency of $\sim 3$ GHz. The observed $\log{R_{\rm X}} \sim -3.6$ confirms that \source is radio loud: hence, the unobscured X-ray component arising below few keV could be attributable to the contribution by a relativistic jet. However, the number of collected photons is too small to have a robust detection of a putative jet component. We note that, based on the observed radio loudness and the relation presented by \citet{miller11}, the X-ray excess due to the jet contribution should be around $\sim 50\%$, albeit with a large uncertainty.

\section{Bolometric Luminosity}\label{sec:lbol}

The total (bolometric) luminosity is one of the most important quantities that describe an AGN, being a measurement of the power emitted from the system and directly connecting it to the accretion mode of the central engine, once the black hole mass is estimated.
To have a thorough view of the accretion state of our GRG, in this Section we derive its bolometric luminosity through different, widely used tracers, and combine them with an estimate of the black hole mass to derive the accretion rate in Eddington units.

\subsection{From X-ray Spectroscopy}

It is well known that, in radio-quiet AGNs, the X-ray luminosity is tightly correlated with its bolometric luminosity \citep[e.g.,][]{kelly08}. Thus, the total luminosity can be easily estimated through a (luminosity-dependent) bolometric correction, $k_{\rm Bol}$. The situation is less clear for radio-loud AGNs, as the jet can potentially contribute to the X-ray emission. As shown earlier, in our case the quality of the \chandra spectrum is not high enough to show any evidence for a jet component, and the \chandra's PSF at the position of the source is too large to disentangle a nuclear component from any kpc-scale extended emission.
Thus, we simply multiply the measured X-ray luminosity by a bolometric correction to obtain the bolometric luminosity \footnote{As explained in \S \ref{subsec:rljet}, an excess-corrected bolometric luminosity can be estimated lowering the final result by $\sim 50\%$.}. \citet{duras20} provide an analytical formula to compute $k_{\rm Bol}$ from $L_{\rm X}$. The significant intrinsic scatter of the relation (0.37 dex) has to be taken into account, together with the uncertainties over the parameters of the relation and those on the X-ray luminosity itself. To do this, we randomly draw the parameters of the correlation and of the X-ray luminosity assuming a Gaussian probability density function for each of them with a dispersion given by their uncertainties, and repeat this procedure $10^4$ times to build a distribution of bolometric corrections keeping into account also the intrinsic scatter. This returns $k_{\rm Bol} = 19.0^{+25.5}_{-10.8}$. The final, X-ray derived bolometric luminosity is $L_{\rm Bol,X} = k_{\rm Bol}L_{\rm X} = 2.5^{+3.4}_{-1.7} \times 10^{45}$ \lumcgs.

\subsection{From Radio Lobes}

Following \citet{ursini18} and \citet{bassani20}, we use the correlation reported by \citet{vanvelzen15} between the 1.4 GHz luminosity of the lobes and the bolometric luminosity of the AGN. The intrinsic scatter of the correlation is 0.47 dex, which is way larger than the uncertainty over the lobes luminosity at 1.4 GHz. Hence, the final lobes-derived bolometric luminosity is $L_{\rm Bol,R} = 4.1^{+8.0}_{-2.7} \times 10^{44}$ \lumcgs.

\subsection{From SED fitting} \label{subsec:sed}

\begin{figure*}
   \centering
   \includegraphics[width=0.8\textwidth]{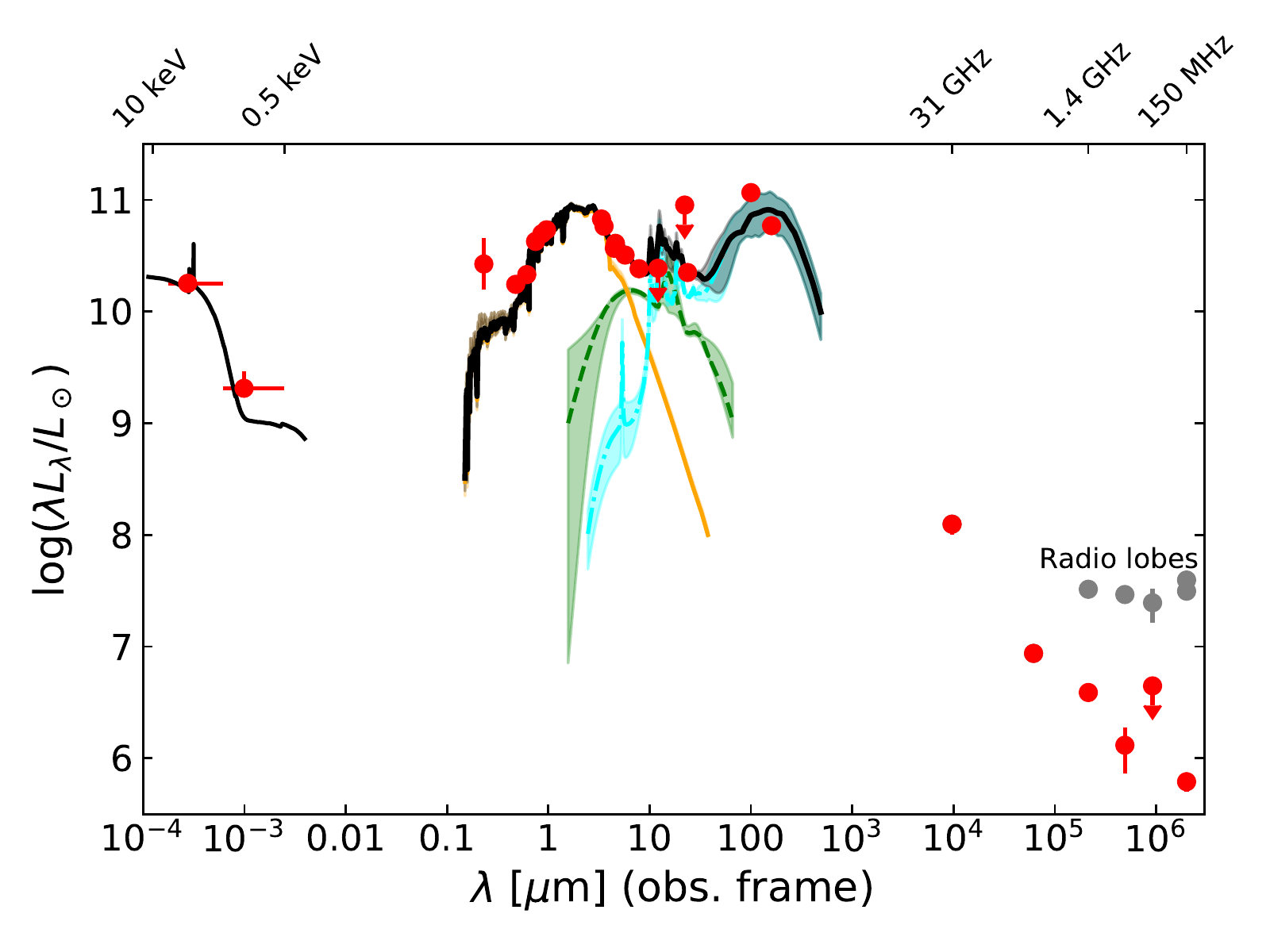}
   \caption{Broadband SED of \source (red data points), covering ten orders of magnitude in wavelength. The leftmost points are the $2-7$ keV and $0.5-2$ keV fluxes from \chandra, respectively, as fitted by the best fit MYTorus model. The range of data points from $\sim 0.1 - 100 ~\mu$m have been fit with SED3FIT 20 times, and the median fit is shown in black, with the colored strips labeling the $68\%$ margins of the distributions for all the components. The host galaxy component of the fit is shown in orange, the AGN torus contribution is in dashed green, and the star formation component is in cyan. The SED is typical of an obscured AGN, with the optical light dominated by the host galaxy. Both a reliable measure of the stellar mass of the host and the bolometric luminosity of the AGN can be inferred from the fit of this portion of the SED. The rightmost data points show in red the core radio SED, while the gray points encompass the lobes emission.}
   \label{fig:sed}
\end{figure*}

Another widely used method to estimate the bolometric luminosity and host stellar mass of an AGN is to perform a fit of its optical-infrared spectral energy distribution (SED), to disentangle the emission of the host galaxy from the emission of the AGN. We thus collected the available UV, optical and IR photometry for \source, for a total of 17 photometric data points (reported in Table \ref{tab:photometry}), and fed them to the SED fitting package SED3FIT \citep{berta13}. SED3FIT is based on MAGPHYS \citep{dacunha08} but includes an AGN template based on the models of \citet{fritz06} and \citet{feltre12}. The median of 20 iterations, with its $1\sigma$ spread, is shown in Figure \ref{fig:sed}. While the cold gas component has a large spread, the host galaxy and AGN contributions (above a few microns) are robustly fit. The stellar mass of the host galaxy is found to be $M_* = 1.17^{+0.09}_{-0.07} \times 10^{11} M_\odot$, and the bolometric luminosity of the AGN is $L_{\rm Bol, SED} = 2.79^{+0.03}_{-0.04} \times 10^{44}$ \lumcgs.

\begin{table}
\caption{Photometry used for the SED fitting.}
\label{tab:photometry}
\centering                                    
\begin{tabular}{l r}          
\hline\hline                        
Telescope/Band & Flux (Jy) \\  
\hline                                   
    {\it Galex}/NUV & $4.69\pm1.94 \times 10^{-6}$\\
    {\it PanSTARRS}/g & $6.37\pm1.02 \times 10^{-6}$  \\
    {\it PanSTARRS}/r & $10.0\pm0.4 \times 10^{-6}$  \\
    {\it PanSTARRS}/i & $24.2\pm0.4 \times 10^{-6}$  \\
    {\it PanSTARRS}/z & $32.8\pm1.1 \times 10^{-6}$  \\
    {\it PanSTARRS}/y & $39.4\pm1.8 \times 10^{-6}$ \\
    {\it Spitzer}/IRAC Ch1 & $157\pm1 \times 10^{-6}$ \\
    {\it Spitzer}/IRAC Ch2 & $126\pm1 \times 10^{-6}$ \\
    {\it Spitzer}/IRAC Ch3 & $140\pm6 \times 10^{-6}$ \\
    {\it Spitzer}/IRAC Ch4 & $145\pm7 \times 10^{-6}$ \\
    {\it Spitzer}/MIPS 24$\mu m$ & $404\pm10 \times 10^{-6}$ \\
    {\it WISE}/W1 & $172\pm6 \times 10^{-6}$ \\
    {\it WISE}/W2 & $143\pm10 \times 10^{-6}$ \\
    {\it WISE}/W3 & $< 225\times 10^{-6}$ \\
    {\it WISE}/W4 & $< 1.52\times 10^{-3}$ \\
    {\it Herschel}/PACS 100$\mu m$ & $8.85\pm11.13 \times 10^{-3}$ \\
    {\it Herschel}/PACS 160$\mu m$ & $7.16\pm13.68 \times 10^{-3}$ \\
\hline                                             
\end{tabular}
\end{table}

\subsection{Bolometric luminosity and accretion rate}

The different methods discussed above return fairly consistent results for the bolometric luminosity. In agreement with previous results for hard X-ray selected GRGs \citep[e.g.,][]{ursini18}, the X-ray-derived bolometric luminosity is larger than the ones derived with other indicators. However, the  significant intrinsic scatter of the relation used to obtain the X-ray bolometric correction makes the X-ray, radio, and SED-derived bolometric luminosities all statistically consistent with each other within 1$\sigma$. Even if the SED fitting returns a bolometric luminosity is significantly lower than the other estimates, and this could be due to an underestimation of the absorption of the AGN emission, it also provides a well-constrained host galaxy stellar mass, from which a black hole mass can be inferred using the well-known scaling relation between the stellar (bulge) and black hole masses \citep{haeringrix04}: $M_{\rm BH} = 1.9^{+0.5}_{-0.4} \times 10^{8} M_\odot$. Of course, we are implicitly assuming that the host galaxy is a spheroid; this assumption is justified, given its brightness and optical colors, typical of elliptical galaxies \citep{tempel11}, and that the majority of radio loud AGNs are hosted by massive, elliptical galaxies \citep[e.g.,][]{best05}. As a final note on luminosity, the poor signal to noise of the AGES optical spectrum of \source prevents to have more insights about the AGN bolometric luminosity from the [OIII] emission line.
\par The Eddington ratio, which is often used as a proxy for the accretion rate onto the black hole, can be computed combining the previously derived bolometric luminosities with the black hole mass, as $\lambda_{\rm Edd} = L_{\rm Bol}/1.26\times10^{38}(M_{\rm BH}/M_\odot)$. Despite the derived Eddington ratios being highly uncertain, they all point to a source accreting at $\sim 1-10\%$ of the Eddington luminosity. Once again, \source behaves like the other hard X-ray selected GRGs, which are powered by radiatively efficient accretion \citep{ursini18}, the only difference being the absence of convincing evidence of a recently restarted radio activity. Table \ref{tab:info} reports a summary of the properties of \source discussed so far.

\section{Large Scale Environment}\label{sec:environment}

Could the environment around GRGs favour their exceptional growth? This question is starting to be systematically explored, although no clear and definitive answer exists at the moment. Early on, it was suggested that GRGs may live in underdense regions, where the circum-galacitc medium density is lower and the jets can expand more freely, resulting in a larger size \citep[e.g.,][]{mack98}. However, \citet{dabhade20} suggested that a non-negligible fraction ($\sim10\%$) of LOFAR-detected GRGs reside in clusters. Also \citet{tang20} found increasing evidence that a sizeable fraction of GRGs could be associated with the brightest cluster galaxies, hence suggesting that underdense environments are not required to grow GRGs. On a similar note, \citet{lanprochaska20} found no evidence that both the small and large scale environments around GRGs play any role in their growth. 
\par
\begin{figure}
   \centering
   \includegraphics[width=0.45\textwidth]{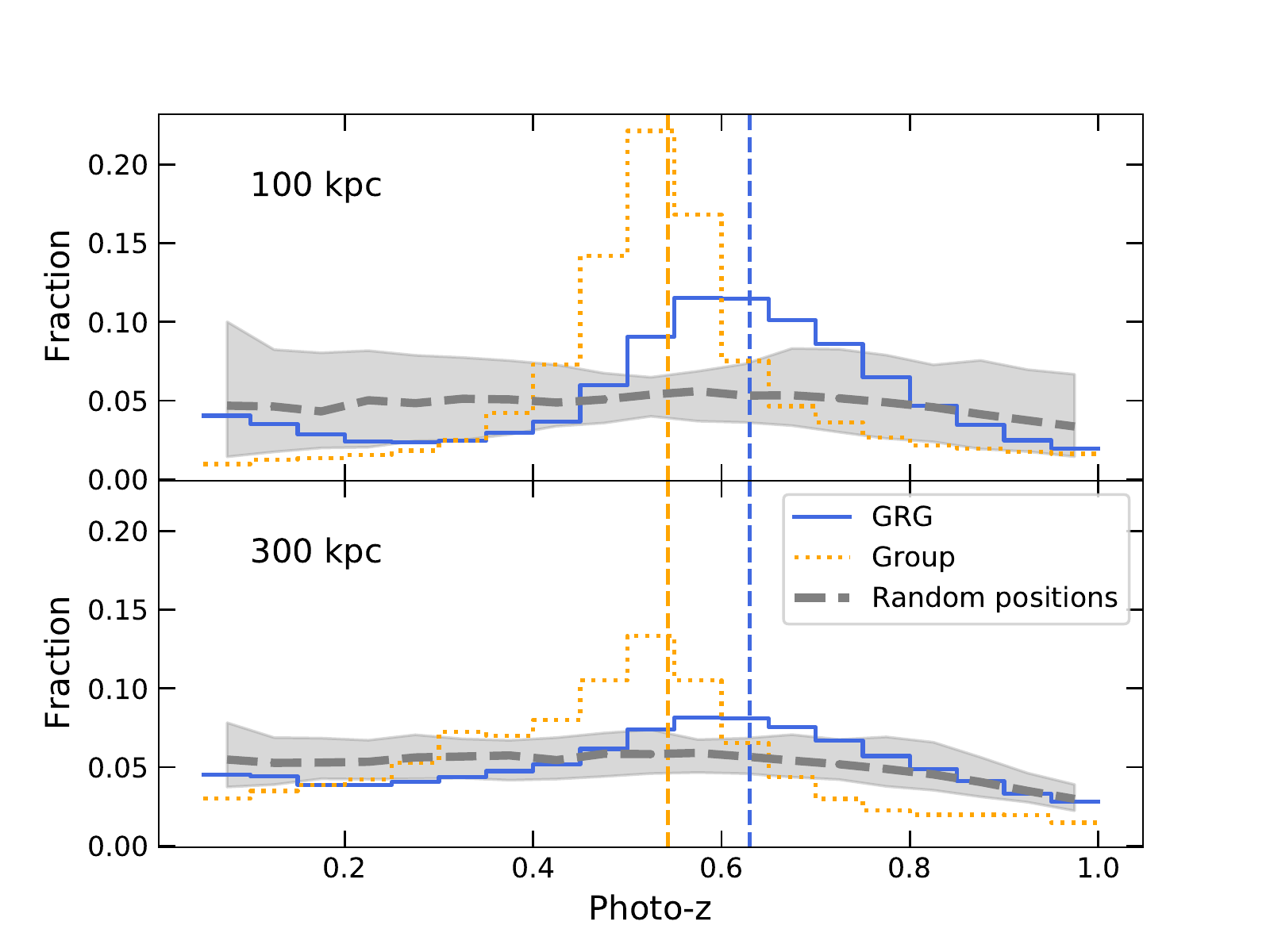}
   \caption{Normalized photometric redshifts distribution, for galaxies with $0<z_{\rm phot}<1$ and within $\sim100$ (top) and $\sim300$ (bottom) projected kpc from \source (blue histogram), for a $z\sim0.54$ X-ray selected group \citep[][XBS41; orange dotted histogram]{vajgel14}, and for 50 random galaxies in the \bootes field with $0.629<z_{\rm phot}<0.631$ (grey dashed line with 68\% spread). The distributions are drawn 5000 times each, taking into account the uncertainties over each photometric redshift as explained in the text. While the peak around the redshift of the GRG is less significant than the group at slightly lower redshift, the distributions of the 50 random positions demonstrate that \source does not live in an underdense environment, which may instead be slightly overdense.}
   \label{fig:env}
\end{figure}

 To investigate the density of \source's nearby environment, we consider all $I$-band detected galaxies \citep[down to a $5\sigma$ limiting AB magnitude of 26.0;][]{brown07} within $15"$ and $44"$ (corresponding to $\sim 100$ and $\sim 300$ projected kpc at $z = 0.629$, respectively) from \source and with a photometric redshift within $0 < z_{\rm phot} < 1$ \citep[see][for details on the photometric redshifts]{duncan18a,duncan18b}. Then, for each redshift satisfying the conditions, we assume a normal probability distribution with a standard deviation approximated by the width of the photo-z posterior ($\sigma = (z_{\rm max}-z_{\rm min})/2$), and pick a random value. We repeat this procedure for 5000 times and build a normalized photometric redshift distribution. The same procedure is applied to a randomly chosen X-ray selected group in the \bootes field, at redshift within $\Delta z = 0.1$ from \source's redshift, namely XBS 41 at $z = 0.54$\footnote{For the group analysis, we adjust the matching radii to $16"$ and $47"$, which translate to projected distances of $\sim 100$ and $\sim 300$ kpc, respectively.} \citep[][]{vajgel14}, and to 50 randomly chosen galaxies in the same \bootes field with $0.629 < z_{\rm phot} < 0.631$. As seen in Figure \ref{fig:env}, while the peak around the redshift of the group XBS 41 is more significant than the peak around \source's redshift, the 50 random galaxies show, on average, flat distributions of their projected companions. These results suggest that \source does not live in an underdense environment.

\section{Discussion}\label{sec:discussion}

We have investigated both the radio and X-ray properties of \source, which resulted consistent with the bulk of the GRG population studied so far in the radio and hard X-ray band, and we demonstrated that its nucleus is actively accreting at moderate rate through a radiatively efficient accretion disk. Moreover, the bolometric luminosity derived from the X-ray spectrum is a factor of six larger than the one derived through the radio lobes - optical luminosity correlation. These results are consistent with what has been found for hard X-ray selected GRGs \citep{ursini18, bassani20}. An extensive multi-wavelength analysis allows to further check whether the the radio core obeys the fundamental plane of black hole activity \citep{merloni03}, which relates the X-ray and core radio luminosity to the black hole mass. Considering the X-ray luminosity and the black hole mass inferred from the host stellar mass (derived from SED fitting), a core radio luminosity at 5 GHz of $\log{(L_{\rm 5, pred}/\text{erg s}^{-1})}\sim40.3$ is predicted, with the observed one being $\log{(L_{\rm 5, obs}/\text{erg s}^{-1})}\sim40.5$. This further supports that the nuclear radio and X-ray activities are connected, and still ongoing. Finally, we have shown that the large scale environment of \source is not underdense with respect to other 50 random galaxies at the same redshift. All these pieces of evidence seem to suggest that \source is rather old and evolved, although with no clear sign of recent restarting of its nuclear activity, unlike what has been shown and found for hard X-ray selected GRGs with deep LOFAR imaging \citep[e.g.,][]{bruni21}.
When compared with the whole GRG population of the SAGAN sample \citep{dabhade20b}, \source is among the highest redshift GRGs known to date, and the highest redshift GRG in the SAGAN sample with a radio spectral index between 150 MHz and 1.4 GHz $\alpha < -1$. On the same line, it is a factor of $4.4$ below the faintest GRG withing the SAGAN sample at the same redshift. Its black hole mass (and conversely, its Eddington ratio) is also among the lowest (highest) of the whole population, but this is not surprising. Indeed, \citet{dabhade20b} estimated the black hole masses using the $M_{\rm BH} - \sigma$ relation for a small subset of GRGs with an available measurement of the bulge velocity dispersion from SDSS optical spectra. As a consequence, the resulting black hole mass distribution is likely skewed to high masses, given that robustly measuring the bulge velocity dispersion is easier, the larger the black hole sphere of influence (and hence its mass). On the other hand, both its radio power $P_{1.4 \text{GHz}} \sim 9 \times 10^{24}$ W Hz$^{-1}$ and jet power, estimated following \citet{dabhade20b} as $Q_{\rm jet}/10^{38}~\text{W} = P_{150}/3\times 10^{27}~\text{W~Hz}^{-1} \sim 3.4\times 10^{43}$ \lumcgs, are perfectly in line with the bulk of the SAGAN sample.
\newline It is important to stress again that the source presented in this paper has been serendipitously detected visually inspecting the superposition of the LOFAR data and the \chandra CDWFS catalog. Medium or soft X-ray selected GRGs, such as the one investigated here, are expected to be more abundant than their hard X-ray selected counterparts: indeed, the hard X-ray selection probes at any redshift the high end of the luminosity function. \source, with the extrapolated $14-195$ keV and $20-100$ keV luminosities of $L_{14-195} \sim 2.2 \times 10^{44}$ \lumcgs and $L_{20-100} \sim 1.3 \times 10^{44}$ \lumcgs, would be a factor of $\sim 30$ and $\sim 80$ below the detection limits of Swift/BAT and INTEGRAL at $z \sim 0.6$, respectively. GRGs are indeed likely much more abundant than thought today\footnote{Furthermore, the definition of GRGs based on projected size $>0.7$ Mpc preferentially selects edge-on systems, missing the GRGs more aligned to our line of sight.} \citep[e.g.,][]{delhaize20}, being the old tail of the population of radio galaxies; a large number of medium X-ray selected GRGs will make up an important control sample to test whether the restarted scenario applies to lower luminosity GRGs as well, and to probe in a comprehensive way their duty cycle\footnote{A control sample of hard X-ray selected, non-giant radio galaxies could provide powerful insight into the restarting view about GRGs as well.}.



\section{Conclusions}\label{sec:conclusions}

In this paper we reported on the serendipitous discovery of a new Giant Radio Galaxy, \source, which lies in the \bootes field at $z=0.629$. Our conclusions are as follows:

\begin{itemize}

\item Its radio properties are in line with the bulk of the GRG population, with the low frequency spectrum dominated by the extended lobes and bright hotspots. There is no obvious indication that \source is a recently restarted GRG like most of the hard X-ray selected ones previously studied in the literature, both from the large scale radio morphology and the core radio spectrum.

\item On the other hand, similarly to other hard X-ray selected GRGs, the nuclear source is bright in X-rays. The X-ray spectrum is moderately obscured, as expected for typical FRII radio galaxies. Furthermore, the nucleus of \source precisely sits on the fundamental plane of black hole activity. The accurate spectral analysis allows to recover an accurate intrinsic X-ray luminosity. No robust detection of a jet component is obtained from the spectral analysis.

\item While we find once again that the X-ray-derived bolometric luminosity is larger than the bolometric luminosity derived from other methods, the uncertainties involved make them all broadly consistent with each other. Both the synchrotron cooling timescale and the position of the source in the P-D diagram suggest the GRG to be a few hundreds Myr old.

\item The number of galaxies around \source seems to favour a slightly overdense environment, further supporting recent findings that underdense environments are not required to grow GRGs.

\item Building a control sample of GRGs detected by soft X-ray observatories is crucial to probe the number density, redshift distribution and restarting behavior displayed by hard X-ray selected GRGs and to fully investigate their duty cycle.

\end{itemize}

All our findings point toward \source being an aged FR II radio galaxy. Future work with ever growing samples of GRGs selected at different frequencies will further elucidate whether there is a real physical difference between GRGs and smaller radio galaxies, or if they simply represent the final stage of their evolution \citep[e.g.,][]{lara04}.

\begin{acknowledgements}
We thank the anonymous referee for a very timely, thorough report that significantly improved the quality of the manuscript.
It is also a pleasure to thank Gabriele Giovannini for fruitful conversations and useful suggestions.
This research made use of LOFAR data products, which were provided by the LOFAR Surveys Key Science project (LSKSP; https://lofar-surveys.org/) and were derived from observations with the International LOFAR Telescope (ILT). LOFAR (van Haarlem et al. 2013) is the Low Frequency Array designed and constructed by ASTRON. It has observing, data processing, and data storage facilities in several countries, which are owned by various parties (each with their own funding sources), and which are collectively operated by the ILT foundation under a joint scientific policy. The efforts of the LSKSP have benefited from funding from the European Research Council, NOVA, NWO, CNRS-INSU, the SURF Co-operative, the UK Science and Technology Funding Council and the Jülich Supercomputing Centre.
E.M. acknowledges financial support from the Czech Science Foundation project No.19-05599Y. This work was supported by the EU-ARC.CZ Large Research Infrastructure grant project LM2018106 of the Ministry of Education, Youth and Sports of the Czech Republic. 

\end{acknowledgements}

\bibliographystyle{aa} 
\bibliography{grg} 

\begin{thebibliography}{63}
\expandafter\ifx\csname natexlab\endcsname\relax\def\natexlab#1{#1}\fi

\bibitem[{{Arnaud}(1996)}]{arnaud96}
{Arnaud}, K.~A. 1996, in Astronomical Society of the Pacific Conference Series,
  Vol. 101, Astronomical Data Analysis Software and Systems V, ed. G.~H.
  {Jacoby} \& J.~{Barnes}, 17

\bibitem[{{Balokovi{\'c}} {et~al.}(2018){Balokovi{\'c}}, {Brightman},
  {Harrison}, {Comastri}, {Ricci}, {Buchner}, {Gandhi}, {Farrah}, \&
  {Stern}}]{balokovic18}
{Balokovi{\'c}}, M., {Brightman}, M., {Harrison}, F.~A., {et~al.} 2018, \apj,
  854, 42

\bibitem[{{Bassani} {et~al.}(2020){Bassani}, {Ursini}, {Malizia}, {Bruni},
  {Panessa}, {Masetti}, {Saviane}, {Monaco}, {Venturi}, {Dallacasa}, {Bazzano},
  \& {Ubertini}}]{bassani20}
{Bassani}, L., {Ursini}, F., {Malizia}, A., {et~al.} 2020, arXiv e-prints,
  arXiv:2010.06427

\bibitem[{{Bassani} {et~al.}(2016){Bassani}, {Venturi}, {Molina}, {Malizia},
  {Dallacasa}, {Panessa}, {Bazzano}, \& {Ubertini}}]{bassani16}
{Bassani}, L., {Venturi}, T., {Molina}, M., {et~al.} 2016, \mnras, 461, 3165

\bibitem[{Belsole {et~al.}(2006)Belsole, Worrall, \& Hardcastle}]{belsole06}
Belsole, E., Worrall, D.~M., \& Hardcastle, M.~J. 2006, Monthly Notices of the
  Royal Astronomical Society, 366, 339–352

\bibitem[{{Berta} {et~al.}(2013){Berta}, {Lutz}, {Santini}, {Wuyts}, {Rosario},
  {Brisbin}, {Cooray}, {Franceschini}, {Gruppioni}, \&
  {Hatziminaoglou}}]{berta13}
{Berta}, S., {Lutz}, D., {Santini}, P., {et~al.} 2013, \aap, 551, A100

\bibitem[{{Best} {et~al.}(2005){Best}, {Kauffmann}, {Heckman}, {Brinchmann},
  {Charlot}, {Ivezi{\'c}}, \& {White}}]{best05}
{Best}, P.~N., {Kauffmann}, G., {Heckman}, T.~M., {et~al.} 2005, \mnras, 362,
  25

\bibitem[{{Boorman} {et~al.}(2016){Boorman}, {Gandhi}, {Alexander}, {Annuar},
  {Ballantyne}, {Bauer}, {Boggs}, {Brandt}, {Brightman}, {Christensen},
  {Craig}, {Farrah}, {Hailey}, {Harrison}, {H{\"o}nig}, {Koss}, {LaMassa},
  {Masini}, {Ricci}, {Risaliti}, {Stern}, \& {Zhang}}]{boorman16}
{Boorman}, P.~G., {Gandhi}, P., {Alexander}, D.~M., {et~al.} 2016, \apj, 833,
  245

\bibitem[{{Brown} {et~al.}(2007){Brown}, {Dey}, {Jannuzi}, {Brand }, {Benson},
  {Brodwin}, {Croton}, \& {Eisenhardt}}]{brown07}
{Brown}, M. J.~I., {Dey}, A., {Jannuzi}, B.~T., {et~al.} 2007, \apj, 654, 858

\bibitem[{{Bruni} {et~al.}(2021){Bruni}, {Brienza}, {Panessa}, {Bassani},
  {Dallacasa}, {Venturi}, {Baldi}, {Botteon}, {Drabent}, {Malizia}, {Massaro},
  {R{\"o}ttgering}, {Ubertini}, {Ursini}, \& {van Weeren}}]{bruni21}
{Bruni}, G., {Brienza}, M., {Panessa}, F., {et~al.} 2021, arXiv e-prints,
  arXiv:2103.00999

\bibitem[{{Bruni} {et~al.}(2020){Bruni}, {Panessa}, {Bassani}, {Dallacasa},
  {Venturi}, {Saripalli}, {Brienza}, {Hern{\'a}ndez-Garc{\'\i}a}, {Chiaraluce},
  {Ursini}, {Bazzano}, {Malizia}, \& {Ubertini}}]{bruni20}
{Bruni}, G., {Panessa}, F., {Bassani}, L., {et~al.} 2020, \mnras, 494, 902

\bibitem[{{Cash}(1979)}]{cash79}
{Cash}, W. 1979, \apj, 228, 939

\bibitem[{{Coppejans} {et~al.}(2016){Coppejans}, {Cseh}, {van Velzen},
  {Falcke}, {Intema}, {Paragi}, {M{\"u}ller}, {Williams}, {Frey}, {Gurvits}, \&
  {K{\"o}rding}}]{coppejans16}
{Coppejans}, R., {Cseh}, D., {van Velzen}, S., {et~al.} 2016, \mnras, 459, 2455

\bibitem[{{Coppejans} {et~al.}(2015){Coppejans}, {Cseh}, {Williams}, {van
  Velzen}, \& {Falcke}}]{coppejans15}
{Coppejans}, R., {Cseh}, D., {Williams}, W.~L., {van Velzen}, S., \& {Falcke},
  H. 2015, \mnras, 450, 1477

\bibitem[{{Croston} {et~al.}(2019){Croston}, {Hardcastle}, {Mingo}, {Best},
  {Sabater}, {Shimwell}, {Williams}, {Duncan}, {R{\"o}ttgering}, {Brienza},
  {G{\"u}rkan}, {Ineson}, {Miley}, {Morabito}, {O'Sullivan}, \&
  {Prandoni}}]{croston19}
{Croston}, J.~H., {Hardcastle}, M.~J., {Mingo}, B., {et~al.} 2019, \aap, 622,
  A10

\bibitem[{{da Cunha} {et~al.}(2008){da Cunha}, {Charlot}, \&
  {Elbaz}}]{dacunha08}
{da Cunha}, E., {Charlot}, S., \& {Elbaz}, D. 2008, \mnras, 388, 1595

\bibitem[{{Dabhade} {et~al.}(2020{\natexlab{a}}){Dabhade}, {Mahato}, {Bagchi},
  {Saikia}, {Combes}, {Sankhyayan}, {R{\"o}ttgering}, {Ho}, {Gaikwad},
  {Raychaudhury}, {Vaidya}, \& {Guiderdoni}}]{dabhade20b}
{Dabhade}, P., {Mahato}, M., {Bagchi}, J., {et~al.} 2020{\natexlab{a}}, \aap,
  642, A153

\bibitem[{{Dabhade} {et~al.}(2020{\natexlab{b}}){Dabhade}, {R{\"o}ttgering},
  {Bagchi}, {Shimwell}, {Hardcastle}, {Sankhyayan}, {Morganti}, {Jamrozy},
  {Shulevski}, \& {Duncan}}]{dabhade20}
{Dabhade}, P., {R{\"o}ttgering}, H.~J.~A., {Bagchi}, J., {et~al.}
  2020{\natexlab{b}}, \aap, 635, A5

\bibitem[{{de Vries} {et~al.}(2002){de Vries}, {Morganti}, {R{\"o}ttgering},
  {Vermeulen}, {van Breugel}, {Rengelink}, \& {Jarvis}}]{devries02}
{de Vries}, W.~H., {Morganti}, R., {R{\"o}ttgering}, H.~J.~A., {et~al.} 2002,
  \aj, 123, 1784

\bibitem[{{Delhaize} {et~al.}(2020){Delhaize}, {Heywood}, {Prescott}, {Jarvis},
  {Delvecchio}, {Whittam}, {White}, {Hardcastle}, {Hale}, {Afonso}, {Ao},
  {Brienza}, {Br{\"u}ggen}, {Collier}, {Daddi}, {Glowacki}, {Maddox},
  {Morabito}, {Prandoni}, {Randriamanakoto}, {Sekhar}, {An}, {Adams}, {Blyth},
  {Bowler}, {Leeuw}, {Marchetti}, {Randriamampandry}, {Thorat}, {Seymour},
  {Smirnov}, {Taylor}, {Tasse}, \& {Vaccari}}]{delhaize20}
{Delhaize}, J., {Heywood}, I., {Prescott}, M., {et~al.} 2020, \mnras
  [\eprint[arXiv]{2012.05759}]

\bibitem[{{Dey} {et~al.}(2019){Dey}, {Schlegel}, {Lang}, {Blum}, {Burleigh},
  {Fan}, {Findlay}, {Finkbeiner}, {Herrera}, {Juneau}, {Landriau}, {Levi},
  {McGreer}, {Meisner}, {Myers}, {Moustakas}, {Nugent}, {Patej}, {Schlafly},
  {Walker}, {Valdes}, {Weaver}, {Y{\`e}che}, {Zou}, {Zhou}, {Abareshi},
  {Abbott}, {Abolfathi}, {Aguilera}, {Alam}, {Allen}, {Alvarez}, {Annis},
  {Ansarinejad}, {Aubert}, {Beechert}, {Bell}, {BenZvi}, {Beutler}, {Bielby},
  {Bolton}, {Brice{\~n}o}, {Buckley-Geer}, {Butler}, {Calamida}, {Carlberg},
  {Carter}, {Casas}, {Castander}, {Choi}, {Comparat}, {Cukanovaite}, {Delubac},
  {DeVries}, {Dey}, {Dhungana}, {Dickinson}, {Ding}, {Donaldson}, {Duan},
  {Duckworth}, {Eftekharzadeh}, {Eisenstein}, {Etourneau}, {Fagrelius},
  {Farihi}, {Fitzpatrick}, {Font-Ribera}, {Fulmer}, {G{\"a}nsicke},
  {Gaztanaga}, {George}, {Gerdes}, {Gontcho}, {Gorgoni}, {Green}, {Guy},
  {Harmer}, {Hernandez}, {Honscheid}, {Huang}, {James}, {Jannuzi}, {Jiang},
  {Joyce}, {Karcher}, {Karkar}, {Kehoe}, {Kneib}, {Kueter-Young}, {Lan},
  {Lauer}, {Le Guillou}, {Le Van Suu}, {Lee}, {Lesser}, {Perreault Levasseur},
  {Li}, {Mann}, {Marshall}, {Mart{\'\i}nez-V{\'a}zquez}, {Martini}, {du Mas des
  Bourboux}, {McManus}, {Meier}, {M{\'e}nard}, {Metcalfe},
  {Mu{\~n}oz-Guti{\'e}rrez}, {Najita}, {Napier}, {Narayan}, {Newman}, {Nie},
  {Nord}, {Norman}, {Olsen}, {Paat}, {Palanque-Delabrouille}, {Peng},
  {Poppett}, {Poremba}, {Prakash}, {Rabinowitz}, {Raichoor}, {Rezaie},
  {Robertson}, {Roe}, {Ross}, {Ross}, {Rudnick}, {Safonova}, {Saha},
  {S{\'a}nchez}, {Savary}, {Schweiker}, {Scott}, {Seo}, {Shan}, {Silva},
  {Slepian}, {Soto}, {Sprayberry}, {Staten}, {Stillman}, {Stupak}, {Summers},
  {Sien Tie}, {Tirado}, {Vargas-Maga{\~n}a}, {Vivas}, {Wechsler}, {Williams},
  {Yang}, {Yang}, {Yapici}, {Zaritsky}, {Zenteno}, {Zhang}, {Zhang}, {Zhou}, \&
  {Zhou}}]{decalsdr8}
{Dey}, A., {Schlegel}, D.~J., {Lang}, D., {et~al.} 2019, \aj, 157, 168

\bibitem[{{Duncan} {et~al.}(2018{\natexlab{a}}){Duncan}, {Brown}, {Williams},
  {Best}, {Buat}, {Burgarella}, {Jarvis}, {Ma{\l}ek}, {Oliver},
  {R{\"o}ttgering}, \& {Smith}}]{duncan18a}
{Duncan}, K.~J., {Brown}, M. J.~I., {Williams}, W.~L., {et~al.}
  2018{\natexlab{a}}, \mnras, 473, 2655

\bibitem[{{Duncan} {et~al.}(2018{\natexlab{b}}){Duncan}, {Jarvis}, {Brown}, \&
  {R{\"o}ttgering}}]{duncan18b}
{Duncan}, K.~J., {Jarvis}, M.~J., {Brown}, M. J.~I., \& {R{\"o}ttgering}, H.
  J.~A. 2018{\natexlab{b}}, \mnras, 477, 5177

\bibitem[{{Duras} {et~al.}(2020){Duras}, {Bongiorno}, {Ricci}, {Piconcelli},
  {Shankar}, {Lusso}, {Bianchi}, {Fiore}, {Maiolino}, {Marconi}, {Onori},
  {Sani}, {Schneider}, {Vignali}, \& {La Franca}}]{duras20}
{Duras}, F., {Bongiorno}, A., {Ricci}, F., {et~al.} 2020, \aap, 636, A73

\bibitem[{{Evans} {et~al.}(2006){Evans}, {Worrall}, {Hardcastle}, {Kraft}, \&
  {Birkinshaw}}]{evans06}
{Evans}, D.~A., {Worrall}, D.~M., {Hardcastle}, M.~J., {Kraft}, R.~P., \&
  {Birkinshaw}, M. 2006, \apj, 642, 96

\bibitem[{{Fanaroff} \& {Riley}(1974)}]{fanaroffriley74}
{Fanaroff}, B.~L. \& {Riley}, J.~M. 1974, \mnras, 167, 31P

\bibitem[{{Feltre} {et~al.}(2012){Feltre}, {Hatziminaoglou}, {Fritz}, \&
  {Franceschini}}]{feltre12}
{Feltre}, A., {Hatziminaoglou}, E., {Fritz}, J., \& {Franceschini}, A. 2012,
  \mnras, 426, 120

\bibitem[{{Fritz} {et~al.}(2006){Fritz}, {Franceschini}, \&
  {Hatziminaoglou}}]{fritz06}
{Fritz}, J., {Franceschini}, A., \& {Hatziminaoglou}, E. 2006, \mnras, 366, 767

\bibitem[{{Fruscione} {et~al.}(2006){Fruscione}, {McDowell}, {Allen},
  {Brickhouse}, {Burke}, {Davis}, {Durham}, {Elvis}, {Galle}, {Harris},
  {Huenemoerder}, {Houck}, {Ishibashi}, {Karovska}, {Nicastro}, {Noble},
  {Nowak}, {Primini}, {Siemiginowska}, {Smith}, \& {Wise}}]{fruscione06}
{Fruscione}, A., {McDowell}, J.~C., {Allen}, G.~E., {et~al.} 2006, in
  \procspie, Vol. 6270, Society of Photo-Optical Instrumentation Engineers
  (SPIE) Conference Series, 62701V

\bibitem[{{Hardcastle}(2018)}]{hardcastle18}
{Hardcastle}, M.~J. 2018, \mnras, 475, 2768

\bibitem[{{H{\"a}ring} \& {Rix}(2004)}]{haeringrix04}
{H{\"a}ring}, N. \& {Rix}, H.-W. 2004, \apjl, 604, L89

\bibitem[{{Jannuzi} \& {Dey}(1999)}]{jannuzi&dey99}
{Jannuzi}, B.~T. \& {Dey}, A. 1999, Astronomical Society of the Pacific
  Conference Series, Vol. 191, {The NOAO Deep Wide-Field Survey}, ed.
  R.~{Weymann}, L.~{Storrie-Lombardi}, M.~{Sawicki}, \& R.~{Brunner}, 111

\bibitem[{{Kalberla} {et~al.}(2005){Kalberla}, {Burton}, {Hartmann}, {Arnal},
  {Bajaja}, {Morras}, \& {P{\"o}ppel}}]{kalberla05}
{Kalberla}, P.~M.~W., {Burton}, W.~B., {Hartmann}, D., {et~al.} 2005, \aap,
  440, 775

\bibitem[{{Kelly} {et~al.}(2008){Kelly}, {Bechtold}, {Trump}, {Vestergaard}, \&
  {Siemiginowska}}]{kelly08}
{Kelly}, B.~C., {Bechtold}, J., {Trump}, J.~R., {Vestergaard}, M., \&
  {Siemiginowska}, A. 2008, \apjs, 176, 355

\bibitem[{{Kenter} {et~al.}(2005){Kenter}, {Murray}, {Forman}, {Jones},
  {Green}, {Kochanek}, {Vikhlinin}, {Fabricant}, {Fazio}, {Brand}, {Brown},
  {Dey}, {Jannuzi}, {Najita}, {McNamara}, {Shields}, \& {Rieke}}]{kenter05}
{Kenter}, A., {Murray}, S.~S., {Forman}, W.~R., {et~al.} 2005, \apjs, 161, 9

\bibitem[{{Kochanek} {et~al.}(2012){Kochanek}, {Eisenstein}, {Cool},
  {Caldwell}, {Assef}, {Jannuzi}, {Jones}, {Murray}, {Forman}, {Dey}, {Brown},
  {Eisenhardt}, {Gonzalez}, {Green}, \& {Stern}}]{kochanek12}
{Kochanek}, C.~S., {Eisenstein}, D.~J., {Cool}, R.~J., {et~al.} 2012, \apjs,
  200, 8

\bibitem[{{Lan} \& {Prochaska}(2020)}]{lanprochaska20}
{Lan}, T.-W. \& {Prochaska}, J.~X. 2020, arXiv e-prints, arXiv:2009.04482

\bibitem[{{Lara} {et~al.}(2004){Lara}, {Giovannini}, {Cotton}, {Feretti},
  {Marcaide}, {M{\'a}rquez}, \& {Venturi}}]{lara04}
{Lara}, L., {Giovannini}, G., {Cotton}, W.~D., {et~al.} 2004, \aap, 421, 899

\bibitem[{{Mack} {et~al.}(1998){Mack}, {Klein}, {O'Dea}, {Willis}, \&
  {Saripalli}}]{mack98}
{Mack}, K.~H., {Klein}, U., {O'Dea}, C.~P., {Willis}, A.~G., \& {Saripalli}, L.
  1998, \aap, 329, 431

\bibitem[{{Magdziarz} \& {Zdziarski}(1995)}]{pexrav95}
{Magdziarz}, P. \& {Zdziarski}, A.~A. 1995, \mnras, 273, 837

\bibitem[{{Masini} {et~al.}(2019){Masini}, {Comastri}, {Hickox}, {Koss},
  {Civano}, {Brigthman}, {Brusa}, \& {Lanzuisi}}]{masini19}
{Masini}, A., {Comastri}, A., {Hickox}, R.~C., {et~al.} 2019, \apj, 882, 83

\bibitem[{{Masini} {et~al.}(2020){Masini}, {Hickox}, {Carroll}, {Aird},
  {Alexander}, {Assef}, {Bower}, {Brodwin}, {Brown}, {Chatterjee}, {Chen},
  {Dey}, {DiPompeo}, {Duncan}, {Eisenhardt}, {Forman}, {Gonzalez}, {Goulding},
  {Hainline}, {Jannuzi}, {Jones}, {Kochanek}, {Kraft}, {Lee}, {Miller},
  {Mullaney}, {Myers}, {Ptak}, {Stanford}, {Stern}, {Vikhlinin}, {Wake}, \&
  {Murray}}]{masini20}
{Masini}, A., {Hickox}, R.~C., {Carroll}, C.~M., {et~al.} 2020, \apjs, 251, 2

\bibitem[{{Massaro} {et~al.}(2020){Massaro}, {Capetti}, {Paggi}, {Baldi},
  {Tramacere}, {Pillitteri}, {Campana}, {Jimenez-Gallardo}, \&
  {Missaglia}}]{massaro20}
{Massaro}, F., {Capetti}, A., {Paggi}, A., {et~al.} 2020, \apjs, 247, 71

\bibitem[{{Merloni} {et~al.}(2003){Merloni}, {Heinz}, \& {di
  Matteo}}]{merloni03}
{Merloni}, A., {Heinz}, S., \& {di Matteo}, T. 2003, \mnras, 345, 1057

\bibitem[{{Miller} {et~al.}(2011){Miller}, {Brandt}, {Schneider}, {Gibson},
  {Steffen}, \& {Wu}}]{miller11}
{Miller}, B.~P., {Brandt}, W.~N., {Schneider}, D.~P., {et~al.} 2011, \apj, 726,
  20

\bibitem[{{Moravec} {et~al.}(2020){Moravec}, {Gonzalez}, {Stern}, {Clarke},
  {Brodwin}, {Decker}, {Eisenhardt}, {Mo}, {Pope}, {Stanford}, \&
  {Wylezalek}}]{moravec20}
{Moravec}, E., {Gonzalez}, A.~H., {Stern}, D., {et~al.} 2020, \apj, 888, 74

\bibitem[{{Muchovej} {et~al.}(2010){Muchovej}, {Leitch}, {Carlstrom},
  {Culverhouse}, {Greer}, {Hawkins}, {Hennessy}, {Joy}, {Lamb}, {Loh},
  {Marrone}, {Miller}, {Mroczkowski}, {Pryke}, {Sharp}, \&
  {Woody}}]{muchovej10}
{Muchovej}, S., {Leitch}, E., {Carlstrom}, J.~E., {et~al.} 2010, \apj, 716, 521

\bibitem[{{Murphy} \& {Yaqoob}(2009)}]{murphy09}
{Murphy}, K.~D. \& {Yaqoob}, T. 2009, \mnras, 397, 1549

\bibitem[{{Murray} {et~al.}(2005){Murray}, {Kenter}, {Forman}, {Jones},
  {Green}, {Kochanek}, {Vikhlinin}, {Fabricant}, {Fazio}, {Brand}, {Brown},
  {Dey}, {Jannuzi}, {Najita}, {McNamara}, {Shields}, \& {Rieke}}]{murray05}
{Murray}, S.~S., {Kenter}, A., {Forman}, W.~R., {et~al.} 2005, \apjs, 161, 1

\bibitem[{{Retana-Montenegro} {et~al.}(2018){Retana-Montenegro},
  {R{\"o}ttgering}, {Shimwell}, {van Weeren}, {Prandoni}, {Brunetti}, {Best},
  \& {Br{\"u}ggen}}]{retanamontenegro18}
{Retana-Montenegro}, E., {R{\"o}ttgering}, H.~J.~A., {Shimwell}, T.~W.,
  {et~al.} 2018, \aap, 620, A74

\bibitem[{{Ricci} {et~al.}(2017){Ricci}, {Trakhtenbrot}, {Koss}, {Ueda}, {Del
  Vecchio}, {Treister}, {Schawinski}, {Paltani}, {Oh}, {Lamperti}, {Berney},
  {Gandhi}, {Ichikawa}, {Bauer}, {Ho}, {Asmus}, {Beckmann}, {Soldi},
  {Balokovi{\'c}}, {Gehrels}, \& {Markwardt}}]{ricci17BASS}
{Ricci}, C., {Trakhtenbrot}, B., {Koss}, M.~J., {et~al.} 2017, \apjs, 233, 17

\bibitem[{{Shimwell} {et~al.}(2017){Shimwell}, {R{\"o}ttgering}, {Best},
  {Williams}, {Dijkema}, {de Gasperin}, {Hardcastle}, {Heald}, {Hoang},
  {Horneffer}, {Intema}, {Mahony}, {Mandal}, {Mechev}, {Morabito}, {Oonk},
  {Rafferty}, {Retana-Montenegro}, {Sabater}, {Tasse}, {van Weeren},
  {Br{\"u}ggen}, {Brunetti}, {Chy{\.z}y}, {Conway}, {Haverkorn}, {Jackson},
  {Jarvis}, {McKean}, {Miley}, {Morganti}, {White}, {Wise}, {van Bemmel},
  {Beck}, {Brienza}, {Bonafede}, {Calistro Rivera}, {Cassano}, {Clarke},
  {Cseh}, {Deller}, {Drabent}, {van Driel}, {Engels}, {Falcke}, {Ferrari},
  {Fr{\"o}hlich}, {Garrett}, {Harwood}, {Heesen}, {Hoeft}, {Horellou},
  {Israel}, {Kapi{\'n}ska}, {Kunert-Bajraszewska}, {McKay}, {Mohan},
  {Orr{\'u}}, {Pizzo}, {Prandoni}, {Schwarz}, {Shulevski}, {Sipior}, {Smith},
  {Sridhar}, {Steinmetz}, {Stroe}, {Varenius}, {van der Werf}, {Zensus}, \&
  {Zwart}}]{shimwell17}
{Shimwell}, T.~W., {R{\"o}ttgering}, H.~J.~A., {Best}, P.~N., {et~al.} 2017,
  \aap, 598, A104

\bibitem[{{Tang} {et~al.}(2020){Tang}, {Scaife}, {Wong}, {Kapi{\'n}ska},
  {Rudnick}, {Shabala}, {Seymour}, \& {Norris}}]{tang20}
{Tang}, H., {Scaife}, A.~M.~M., {Wong}, O.~I., {et~al.} 2020, \mnras, 499, 68

\bibitem[{{Tasse} {et~al.}(2020){Tasse}, {Shimwell}, {Hardcastle},
  {O'Sullivan}, {van Weeren}, {Best}, {Bester}, {Hugo}, {Smirnov}, {Sabater},
  {Calistro-Rivera}, {de Gasperin}, {Morabito}, {R{\"o}ttgering}, {Williams},
  {Bonato}, {Bondi}, {Botteon}, {Br{\"u}ggen}, {Brunetti}, {Chy{\.z}y},
  {Garrett}, {G{\"u}rkan}, {Jarvis}, {Kondapally}, {Mandal}, {Prandoni},
  {Repetti}, {Retana-Montenegro}, {Schwarz}, {Shulevski}, \& {Wiaux}}]{tasse20}
{Tasse}, C., {Shimwell}, T., {Hardcastle}, M.~J., {et~al.} 2020, arXiv
  e-prints, arXiv:2011.08328

\bibitem[{{Tempel} {et~al.}(2011){Tempel}, {Saar}, {Liivam{\"a}gi}, {Tamm},
  {Einasto}, {Einasto}, \& {M{\"u}ller}}]{tempel11}
{Tempel}, E., {Saar}, E., {Liivam{\"a}gi}, L.~J., {et~al.} 2011, \aap, 529, A53

\bibitem[{{Terashima} \& {Wilson}(2003)}]{terashimawilson03}
{Terashima}, Y. \& {Wilson}, A.~S. 2003, \apj, 583, 145

\bibitem[{{Ursini} {et~al.}(2018){Ursini}, {Bassani}, {Panessa}, {Bird},
  {Bruni}, {Fiocchi}, {Malizia}, {Saripalli}, \& {Ubertini}}]{ursini18}
{Ursini}, F., {Bassani}, L., {Panessa}, F., {et~al.} 2018, \mnras, 481, 4250

\bibitem[{{Vajgel} {et~al.}(2014){Vajgel}, {Jones}, {Lopes}, {Forman},
  {Murray}, {Goulding}, \& {Andrade-Santos}}]{vajgel14}
{Vajgel}, B., {Jones}, C., {Lopes}, P. A.~A., {et~al.} 2014, \apj, 794, 88

\bibitem[{{van Haarlem} {et~al.}(2013){van Haarlem}, {Wise}, {Gunst}, {Heald},
  {McKean}, {Hessels}, {de Bruyn}, {Nijboer}, {Swinbank}, {Fallows},
  {Brentjens}, {Nelles}, {Beck}, {Falcke}, {Fender}, {H{\"o}randel},
  {Koopmans}, {Mann}, {Miley}, {R{\"o}ttgering}, {Stappers}, {Wijers},
  {Zaroubi}, {van den Akker}, {Alexov}, {Anderson}, {Anderson}, {van Ardenne},
  {Arts}, {Asgekar}, {Avruch}, {Batejat}, {B{\"a}hren}, {Bell}, {Bell}, {van
  Bemmel}, {Bennema}, {Bentum}, {Bernardi}, {Best}, {B{\^\i}rzan}, {Bonafede},
  {Boonstra}, {Braun}, {Bregman}, {Breitling}, {van de Brink}, {Broderick},
  {Broekema}, {Brouw}, {Br{\"u}ggen}, {Butcher}, {van Cappellen}, {Ciardi},
  {Coenen}, {Conway}, {Coolen}, {Corstanje}, {Damstra}, {Davies}, {Deller},
  {Dettmar}, {van Diepen}, {Dijkstra}, {Donker}, {Doorduin}, {Dromer}, {Drost},
  {van Duin}, {Eisl{\"o}ffel}, {van Enst}, {Ferrari}, {Frieswijk}, {Gankema},
  {Garrett}, {de Gasperin}, {Gerbers}, {de Geus}, {Grie{\ss}meier}, {Grit},
  {Gruppen}, {Hamaker}, {Hassall}, {Hoeft}, {Holties}, {Horneffer}, {van der
  Horst}, {van Houwelingen}, {Huijgen}, {Iacobelli}, {Intema}, {Jackson},
  {Jelic}, {de Jong}, {Juette}, {Kant}, {Karastergiou}, {Koers}, {Kollen},
  {Kondratiev}, {Kooistra}, {Koopman}, {Koster}, {Kuniyoshi}, {Kramer},
  {Kuper}, {Lambropoulos}, {Law}, {van Leeuwen}, {Lemaitre}, {Loose}, {Maat},
  {Macario}, {Markoff}, {Masters}, {McFadden}, {McKay-Bukowski}, {Meijering},
  {Meulman}, {Mevius}, {Middelberg}, {Millenaar}, {Miller-Jones}, {Mohan},
  {Mol}, {Morawietz}, {Morganti}, {Mulcahy}, {Mulder}, {Munk}, {Nieuwenhuis},
  {van Nieuwpoort}, {Noordam}, {Norden}, {Noutsos}, {Offringa}, {Olofsson},
  {Omar}, {Orr{\'u}}, {Overeem}, {Paas}, {Pand ey-Pommier}, {Pandey}, {Pizzo},
  {Polatidis}, {Rafferty}, {Rawlings}, {Reich}, {de Reijer}, {Reitsma},
  {Renting}, {Riemers}, {Rol}, {Romein}, {Roosjen}, {Ruiter}, {Scaife}, {van
  der Schaaf}, {Scheers}, {Schellart}, {Schoenmakers}, {Schoonderbeek},
  {Serylak}, {Shulevski}, {Sluman}, {Smirnov}, {Sobey}, {Spreeuw}, {Steinmetz},
  {Sterks}, {Stiepel}, {Stuurwold}, {Tagger}, {Tang}, {Tasse}, {Thomas},
  {Thoudam}, {Toribio}, {van der Tol}, {Usov}, {van Veelen}, {van der Veen},
  {ter Veen}, {Verbiest}, {Vermeulen}, {Vermaas}, {Vocks}, {Vogt}, {de Vos},
  {van der Wal}, {van Weeren}, {Weggemans}, {Weltevrede}, {White}, {Wijnholds},
  {Wilhelmsson}, {Wucknitz}, {Yatawatta}, {Zarka}, {Zensus}, \& {van
  Zwieten}}]{vanhaarlem13}
{van Haarlem}, M.~P., {Wise}, M.~W., {Gunst}, A.~W., {et~al.} 2013, \aap, 556,
  A2

\bibitem[{{van Velzen} {et~al.}(2015){van Velzen}, {Falcke}, \&
  {K{\"o}rding}}]{vanvelzen15}
{van Velzen}, S., {Falcke}, H., \& {K{\"o}rding}, E. 2015, \mnras, 446, 2985

\bibitem[{{Williams} {et~al.}(2013){Williams}, {Intema}, \&
  {R{\"o}ttgering}}]{williams13}
{Williams}, W.~L., {Intema}, H.~T., \& {R{\"o}ttgering}, H.~J.~A. 2013, \aap,
  549, A55

\bibitem[{{Williams} {et~al.}(2016){Williams}, {van Weeren}, {R{\"o}ttgering},
  {Best}, {Dijkema}, {de Gasperin}, {Hardcastle}, {Heald}, {Prandoni},
  {Sabater}, {Shimwell}, {Tasse}, {van Bemmel}, {Br{\"u}ggen}, {Brunetti},
  {Conway}, {En{\ss}lin}, {Engels}, {Falcke}, {Ferrari}, {Haverkorn},
  {Jackson}, {Jarvis}, {Kapi{\'n}ska}, {Mahony}, {Miley}, {Morabito},
  {Morganti}, {Orr{\'u}}, {Retana-Montenegro}, {Sridhar}, {Toribio}, {White},
  {Wise}, \& {Zwart}}]{williams16}
{Williams}, W.~L., {van Weeren}, R.~J., {R{\"o}ttgering}, H.~J.~A., {et~al.}
  2016, \mnras, 460, 2385

\bibitem[{{Willis} {et~al.}(1974){Willis}, {Strom}, \& {Wilson}}]{willis74}
{Willis}, A.~G., {Strom}, R.~G., \& {Wilson}, A.~S. 1974, \nat, 250, 625

\end{thebibliography}

\end{document}